\documentclass[onecolumn,pdflatex,nofootinbib,superscriptaddress,floatfix]{revtex4-1}
\usepackage{graphicx}
\usepackage{bm}
\usepackage{placeins}
\usepackage{xspace}
\usepackage{bm,amsmath,amssymb,amsfonts,gensymb}
\usepackage{comment}
\usepackage{hyperref}
\hypersetup{
    colorlinks=true,
    linkcolor=blue,
    citecolor=blue,
    filecolor=blue,      
    urlcolor=blue,
}
\usepackage{cleveref}

\usepackage[english]{babel}
\newcommand{\eq}[1]{\begin{align} #1 \end{align}}

\newcommand{\be}{\begin{equation}}
\newcommand{\ee}{\end{equation}}

\usepackage{color}

\usepackage[english]{babel}
\usepackage{verbatim}
\begin{document}

\title{
Quasi-normal modes of naked singularities in presence of  non-linear scalar fields
 }
\author{O.~S.~Stashko}
\affiliation{Princeton University, Princeton, NJ 08544}
\affiliation{Goethe Universität, Max-von-Laue Str. 1, Frankfurt am Main, 60438, Germany}

\author{O.~V.~Savchuk}
\affiliation{Facility for Rare Isotope Beams, Michigan State University, East Lansing, MI 48824 USA}
\affiliation{Bogolyubov Institute for Theoretical Physics, 03680 Kyiv, Ukraine}

\author{V.~I.~Zhdanov}
\affiliation{Taras Shevchenko National University of Kyiv, Ukraine}
\affiliation{Igor Sikorsky Kyiv Polytechnic Institute, Ukraine}

\date{\today}

\pacs{11}


\begin{abstract}
We study  linear perturbations against  static spherically symmetric  background configurations of General Relativity with a   real  scalar field (SF), which is minimally coupled  with gravity; it is  non-linear due to the presence of the self-action  potential. The background solutions  have a naked singularity at the center of the configuration. The focus is on  the stability of the  background and fundamental frequencies of the  quasi-normal modes (QNM) of the axial perturbations in the Regge-Wheeler gauge. The problem is reduced to one hyperbolic master  equation with an effective  potential $W_{\rm eff}$, which  turns out to be positive for a general non-negative SF potential; this ensures the linear stability with respect to this kind of perturbations.

For  numerical simulations, the SF potential was chosen in the power-law form  $V(\phi)\sim\phi^{2n}$ with $2<n\le 40$.    We extracted   the fundamental frequencies of  QNM  for different $n$ and various sets of the  background configuration parameters. The results show that even for a small background SF, there is a significant   difference between the    fundamental  frequencies and ones in case of  the Schwarzschild background. The  results are also compared with the case of the Fisher-Janis-Newman-Winicour background dealing with a massless linear scalar field. 
\end{abstract}

\maketitle

\section*{Introduction}
Successful detections of gravitational waves    \cite{GW150914,catalog-GW}  and the emerging prospects for multi-messenger astronomy \cite{MultimessengerAstr} have opened up new possibilities to test the gravitational theories in the strong field regime. This stimulated a surge of attention to  damped oscillations of perturbed astrophysical objects as a source of gravitational radiation. These oscillations, known as quasi-normal modes  \cite{Nollert1999,Kokkotas1999LRR, Berti2009, Konoplya2011} carry important information about properties of relativistic objects, which, in particular, can be used, for example, to distinguish the black holes from the naked singularities arising in various gravitational theories. In this paper, we analyze theories with minimally coupled self-interacting scalar field (SF), which may be considered as simplest modifications  of  the General Relativity. On the other hand, self-interacting SF is a workhorse of huge number of models dealing with the inflation of the Universe,  dark energy etc. 

If  static spherically symmetric configuration  contains only gravitational and scalar fields satisfying the conditions of asymptotic flatness, then the appearance of a naked singularity (NS) is expected in this system.  An example is the well-known Fisher-Janis-Newman-Winicour (FJNW) solution \cite{Fisher,JNW,Virbhadra_1997}. 
The existence of NS has been demonstrated in case of a spherically symmetric static configuration with a fairly general SF potential \cite{ZhdSt,Strongly2021}. 
Here we are faced with the main question:   whether there are static configurations with SF and NS in our Universe at all. 
The no-hair theorems \cite{Bekenstein1972,Bekenstein1972a} say that the black hole with a  regular non-zero SF  cannot exist; so one can suppose that the scalar hair must simply dissipate  during the collapse into the black hole. This would be   in accordance  with the Cosmic Censorship hypothesis \cite{Penrose1965}. 
 On the other hand, there exist solutions in the General Relativity and the other theories that describe NS  (see, e.g., \cite{Christodoulou1984,Ori1987,Joshi1993,Joshi_2013,Ong2020}), though, how they  can be formed in our Universe is still a moot point. In this regard,  it is appropriate to look for observational effects that can distinguish between normal black holes and their mimickers that contain NS.

The quasinormal modes (QNM) of the space-time vibrations emerging around relativistic objects  can be good probes of different gravitational theories. Much  attention has been paid to perturbations of the black holes and regular compact astrophysical objects (see, e.g., reviews \cite{Kokkotas1999LRR,Berti2009, Konoplya2011}).
QNM near NS have been studied in  \cite{Chirenti2012, Khamesra2014,Santos2019, Chowdhury2020,Chowdhury2022}; here, most  works essentially deal  with spherically symmetric  background space-times, where  an analytic representation of a background metric is available, such as in the Reissner-Nordstrom  \cite{Chirenti2012,  Santos2019} and FJNW \cite{Chowdhury2020} cases. 
Obviously, considerations of the linear perturbations are deeply connected with the stability analysis (see, e.g., \cite{Khamesra2014,Chowdhury2022,Dotti_2022} and references therein).

In the present paper, we study axial perturbations  against the background formed by a isolated static spherically symmetric configuration of General Relativity in  presence of a minimally-coupled non-linear real SF. Our findings extend  the results of 
\cite{Chowdhury2020},  which deal   with pure FJNW (massless linear case).  In our case, the nonlinearity is introduced using a general SF-potential $V(\phi)$ satisfying the condition $\phi V'(\phi)\ge 0$, which grows more slowly than exponent for $|\phi|\to\infty$.
In the numerical simulations we are dealing  with the power-law SF  potential. 

In Section \ref{General} the basic relations for static  spherically symmetric background, are presented. 
In Section \ref{monomial_V} we turn to numerical methods for the case of the power-law SF potential. We use the method of backwards integration \cite{ZhdSt,SZA2} to obtain the   solutions of Einstein + SF equations for the background. In Section \ref{QNM-frequencies} we use the time-domain integration    method \cite{Chirenti_2007} to solve the wave equations for the  perturbations. Given these  solutions, we use the Prony method \cite{prony} to extract the frequencies of the fundamental QNM. The final section summarizes the results.

\section{Analytic relations for  general SF potential}\label{General}
The action of standard General Relativity in the presence of a minimally-coupled real SF  $\phi$ is given by\footnote{Units: $c=1$; the metric signature is (+\,-\,-\,-), $R^\alpha_{\,\,\beta\gamma\delta}=-R^\alpha_{\,\,\beta\delta\gamma}=\partial_\gamma\Gamma^\alpha_{\beta\delta}-...\,;\,\,R_{\mu\nu}=R^\alpha_{\,\,\mu\alpha\nu}$. Below we rescale units of mass, length and SF so as to put $G=1$ and to remove a coupling constant in the SF potential.}
\begin{equation}
S=\int  d^4\, x \sqrt{-g}\left(-\frac{R}{16\pi G }+\frac{1}{2}\partial_{\mu}\phi\partial^{\mu}\phi-V(\phi)\right)\,.
\end{equation}

The background metric describes the static spherically symmetric space-time in Schwarzschild-like (curvature) coordinates:
\begin{equation}
ds^2 = e^{\alpha(r)}dt^2 - e^{\beta(r)}dr^2 - r^2 d\Omega^2,
\label{metric}
\end{equation}
where $d\Omega^2=d\theta^2+(\sin\theta)^2 d\varphi^2$ and the radial variable $r>0$.

 In static spherically symmetric case we deal with two independent first-order Einstein equations 
\begin{equation}
\frac{d}{dr}\left[r\left(e^{-\beta}-1\right)\right]=-8\pi r^2[e^{-\beta}\phi'^2/2+V(\phi)],
\label{Ein_1-0}
\end{equation}
\be
 	\label{Ein_2-0}
 	r e^{-\beta}\frac{d\alpha}{dr}+e^{-\beta}-1 =8\pi r^2 [e^{-\beta}\phi'^2/2- V(\phi)],
\ee
and  one second order equation for SF $\phi=\phi(r)$ 
\be
 \label{equation-phi}
\frac{d}{dr}\left[r^2 e^{\frac{\alpha-\beta}{2}}\frac{d\phi}{dr}\right]=r^2 e^{\frac{\alpha+\beta}{2}}V'(\phi)\, .
\ee
The equations (\ref{Ein_1-0}, \ref{Ein_2-0}, \ref{equation-phi}) define the background solutions under appropriate asymptotic conditions. 

A small perturbation of the background solution will be considered in the linear approximation. The space-time metric and SF is then:
\eq{\label{Metric_perturbations}
g_{\mu\nu}=g_{\mu\nu}^{(0)}+h_{\mu\nu},~\phi(x^\mu)=\phi_0(r)+\delta\phi(x^{\mu}), } 
where $g_{\mu\nu}^{(0)}$ and   $\phi_0(r)$   represent the static spherically symmetric solution of (\ref{Ein_1-0}-\ref{equation-phi}), while $h_{\mu\nu}=h_{\mu\nu}(x^{\mu})$ and $\delta\phi(x^{\mu})$ represent the   perturbations. There is extensive literature on the gravitational perturbations against spherically symmetric background  (see, e.g., \cite{Chandrasekhar1998, Berti2009}). 

The perturbation can be separated into axial and polar parts\footnote{In this case ``axial'' and ``polar'' \cite{Berti2009,Chandrasekhar1998}  
corresponds to ``odd'' and ``even'' according to the initial paper \cite{Regge-Wheeler}.}, which can be treated independently. In this paper, we  focus solely on the axial perturbations. 
The metrics of the axial gravitational perturbations $h_{\mu\nu}$ in the Regge-Wheeler gauge \cite{Regge-Wheeler} take the following form
\begin{eqnarray}
h^{axial}_{\mu \nu}= \left[
 \begin{array}{cccc}
 0 & 0 &0 & h_0(t,r)
\\ 0 & 0 &0 & h_1(t,r)
\\ 0 & 0 &0 & 0
\\ h_0(t,r) & h_1(t,r) &0 &0
\end{array}\right]
\left(\sin\theta\frac{\partial}{\partial\theta}\right)
P_{l}(\cos\theta)\,, \label{pert_axial}
\end{eqnarray}
where $h_0(t,r)$ and $h_1(t,r)$ are two unknown functions, and $P_l(x)$  is the Legendre polynomial\footnote{Due to the spherical symmetry of the background configuration, more general case that involves $Y_{lm}$, leads to the same master equation and the same QNM frequencies. The perturbations with $l=1$ can be removed by gauge transformation \cite{Vishveshwara}. } with $l\geq2$. After substitution of  (\ref{pert_axial}) to the Einstein-SF equations and retaining only linear terms, we obtain:

\eq{\label{pert1}
-{h_0^\prime} \left( \alpha '+ \beta '\right)+2
   h_0^{\prime\prime}-{h_0} \left(\frac{2 {l(l+1)}}{r^2} e^{\beta }+2 
   \alpha ''- \alpha ' \beta '+ \alpha
   '^2\right)+\dot{h_1}\left( \alpha '+ \beta '-\frac{4}{r}\right)-2  \dot{h_1^\prime}=0,
}
\eq{\label{pert2}
 \dot{h_0^\prime}-\frac{2}{r}  \dot{h_0}-
   \ddot{h_1}- \frac{l(l+1)-2}{r^2} e^{\alpha} {h_1}=0,
}
\eq{\label{pert3}
2 
   {h_1'}+ {h_1} \left(\alpha'-\beta
  '\right)-2 e^{\beta-\alpha} \dot{h_0}
=0.}
Here $\alpha, \beta$ correspond to the background solution. In agreement with \cite{Kobayashi_odd}, the perturbed part of  SF equals to zero \eq{\delta\phi=0.} 
Equation (\ref{pert1}) is redundant, because for  perturbations $\sim e^{-i\omega t}$ it can be obtained from (\ref{pert2}, \ref{pert3}). 

After substitution 
\eq{
h_1(t,r)=r  e^{(\beta-\alpha)/2}\Psi(t,r), 
}
and introducing 
the  ``tortoise"  coordinate $r^{*}$  
\eq{\label{tortoise}\frac{dr^*}{dr}=e^{(\beta-\alpha)/2} \,,\quad r^*(0)=0\,,}
we can combine (\ref{pert2}, \ref{pert3}) into a single master equation 
\begin{equation} 
\left(
\frac{\partial^2}{\partial t^2}-\frac{\partial^2}{\partial r^*{}^2}\right)\Psi(t,r^*)+W_{\rm eff}(r,l)\Psi(t,r^*)=0,
\label{master_eq}
\end{equation}
where
\begin{equation} 
W_{\rm eff}(r,l)=e^{\alpha-\beta}\left(\frac{\beta'-\alpha'}{2r}+e^{\beta}\frac{(l-1)(l+2)}{r^2}+\frac{2}{r^2}\right),
\label{wave_potential}
\end{equation}
 will be referred to as the ``wave potential''.

Using equations (\ref{Ein_1-0}, \ref{Ein_2-0}), we can rewrite   (\ref{wave_potential}) as
\begin{equation}\label{positive_Veff}
W_{\rm eff}(r^*)\equiv W_{\rm eff}(r,l)=e^\alpha \left[8\pi V(\phi_0(r))  +\frac{(l-1)(l+2)-1}{r^2}  \right]+ \frac{3}{r^2}e^{\alpha-\beta}\,,
\end{equation}
where $r=r(r^*)$ is defined by (\ref{tortoise}). 

Further  we fix $l\ge2$; then $\forall r$: $W_{\rm eff}(r^*)>0$ in case of an arbitrary 
non-negative SF potential. This leads to the absence of   bound  states and stability of solutions of equation (\ref{master_eq}).

    Under the condition of convergence of the integrals involved and for 
\begin{equation}
\lim_{r^*\to 0+0}  
    \left[\frac{\partial  \Psi(t,r^*)}{\partial t}\frac{\partial \Psi(t,r^*)}{\partial r^*}\right]=0 \quad {\rm and}\quad      \lim_{r^*\to \infty}\left[\frac{\partial  \Psi(t,r^*)}{\partial t}\frac{\partial \Psi(t,r^*)}{\partial r^*}\right]=0  \,,
\end{equation}
we have a conserved quantity 
\begin{equation}
E(t,\Psi)=   \int\limits_0^\infty dr^* \left\{ \left( \frac{\partial \Psi}{\partial t}\right)^2+\left( \frac{\partial \Psi}{\partial r^*}\right)^2 + W_{\rm eff}(r^*)\Psi^2\right\}.
\label{conserved_norm}
\end{equation}

So far, we have formally dealt with arbitrary $V(\phi)$. Now, we need more detailed information concerning the properties of the system near the singular center. To do so, we impose the conditions \cite{ZhdSt} 
\eq{\label{eq:cond_V}
V(0)= 0,\quad \phi V'(\phi)\ge 0,\quad  |V'(\phi)|\le C\exp(\kappa|\phi|^{p}),} 
where $C, \kappa, p$ are positive constants, and $p<1$. Under the latter condition,    the terms with $V(\phi)$ in (\ref{Ein_1-0},\ref{Ein_2-0},\ref{equation-phi}) do not directly contribute into the first two orders of the asymptotic expansion 
of $\phi,\alpha,\beta$   for $r\to 0$  yielding
 \be
\phi(r)= -\xi\ln \left(\frac{r}{r_g}\right)+\phi_0 +O(r^{\eta+1}) \,,
\label{asymptotics_NS_phi}
 \ee
\be
\alpha (r)= (\eta -1)\ln \left(\frac{r}{r_g}\right) +\alpha_0+O(r^{\eta+1}), \quad   \beta (r)= (\eta +1)\ln \left(\frac{r}{r_g}\right) +\beta_0+O(r^{\eta+1}),
\label{asymptotics_NS}
\ee
 where $\eta=4\pi\xi^2> 0$, and  the constants $\xi\,,\eta, \alpha_0\,, \beta_0$ and $\phi_0$, which  depend on  $Q,M,n$, have been used to substantiate the asymptotic of $W_{\rm eff}(r^*)$ below. 
The numerical values can be determined for the explicit form of $V(\phi)$.  Note that the the conditions imposed on $V$ are satisfied in case of  the power-law potential considered below.

Taking into account (\ref{asymptotics_NS_phi}, \ref{asymptotics_NS}), we get  asymptotic relations near the singularity ($r\to{0},\, l\ge 2$)
\begin{equation}
r^*=\frac{r^2}{2r_g}e^{(\beta_0-\alpha_0)/2}\left[1+O\left(r^{\eta+1}\right)\right],\quad 
W_{\rm eff}(r,l)= \frac{3e^{\alpha_0-\beta_0}r_g^2}{r^4}\left[1+O\left(r^\eta\right)\right], \quad
W_{\rm eff} (r^* )=\frac{3}{4r^*{}^2}\left[1+O\left({r^*{}^{\eta/2}}\right)\right].
\label{asy-V-zero}
\end{equation}

In the presence of a naked singularity the space-time is not  globally hyperbolic and  evolution governed by (\ref{master_eq}) may be non-unique. However, it was shown \cite{Wald1980,2003CQGra..20.3815I} (see also \cite{Ishibashi-Hosoya1999,Gibbons2005,Dotti_2022}) that we immediately can have a well-defined dynamic, if there is a unique self-adjoint extension $A_E$  of the spatial part of the wave operator (\ref{master_eq}). Using this, one  can specify the boundary conditions at the singularity. We shall show that this is just the case of equation (\ref{master_eq}). 

Equation (\ref{master_eq}) can be written as 
\eq{\label{master_eq_2}
\frac{\partial^2\Psi}{\partial t^2}=-A\Psi\,,\quad A=-\frac{d^2}{dr^*{}^2}+W_{\rm eff}(r^*),\quad r^*\in(0,\infty)}
Following \cite{Wald1980,2003CQGra..20.3815I}, we   consider a positive symmetric operator $A$ defined on $C_0^{\infty}(0,\infty)$ functions (with a compact support). Then we consider a self-adjoint extension $A_E$ to have a well-defined  dynamics governed by (\ref{master_eq}) leading to  boundary conditions    corresponding to $A_E$ \cite{Wald1980,2003CQGra..20.3815I}.
To check the essential self-adjointness we can use the Weyl`s limit point -- limit circle theorem \cite{Reed-Simon1975}.

A continuous function $W(r^*)$,    $r^*\in (0,\infty)$, is said to be in the in the limit point case at zero (or infinity) \cite{Reed-Simon1975}, if for some $\lambda$, equation  
\begin{equation}\label{limit_point_case}
-\frac{d^2\Psi}{dr^*{}^2}+W_{\rm eff}{(r^*)}\Psi=-\lambda^2\Psi ,
\end{equation}
has at least one solution near zero (or infinity), which is not square integrable.

Now we assume that $V(\phi(r(r^*)))$ in (\ref{positive_Veff}) decays faster than $r^*{}^{-3}$ for $r^*{}\to\infty$. Then 
\begin{equation}\label{W_inf}
W_{\rm eff}(r^*)=\frac{l(l+1)}{r^*{}^2}\left[1+O\left(\frac{\ln r^*}{r^*}\right)\right]\,.
\end{equation}
Note that this is fulfilled in the case of the power-law potential  considered in the next section. Then 
 we have asymptotic solutions to (\ref{limit_point_case})
\eq{
\Psi(r^*)\simeq\sqrt{r^*}\left(C_1 J_{l+\frac{1}{2}}(\lambda r^*)+C_2 Y_{l+\frac{1}{2}}(\lambda r^*)\right)\,, \quad r^*\to{\infty}\,,
}
which obviously is  not square-integrable if we choose  $\lambda$ with ${\rm Im}\lambda\ne 0$; i.e. $W_{\rm eff} (r^*)$ is in the limit  point case at $r\to\infty$.

Also, near the singularity  we have
\eq{\Psi(r^*)\simeq C_1 r^*{}^{\frac{3}{2}}+C_2r^*{}^{-\frac{1}{2}}\,, \quad r^*\to 0\,,\label{Psi_near_0}}
where the  solution with $C_2\ne 0$ is not square integrable. Thus $W_{\rm eff} (r^*)$ is in the limit  point case for $r^*\to 0$ as well. 
According to the Weyl limit point -- limit circle theorem \cite{Reed-Simon1975} operator   $A$ defined  is essentially self-adjoint on   $C^\infty_0(0,\infty)$.  

Now $A$ can be extended to a domain $S$, where $S\subset L^{2}$ is a set of functions $f(r^*)$, such that   (i)   $f(0)=0$, (ii) $|f(r^*)|$ and $|df/dr^*|$ decay  fast enough as $r^*\to\infty$, so that the integral    $\int\limits_0^\infty (|f'(x)|^2+W_{\rm eff}(x)|f(x)|^2) dx<\infty$.  
The condition at $x=0$  is  sufficient to extract the unique  solution of (\ref{Psi_near_0}) with $C_2=0$ and to justify the null Dirichlet boundary condition at the center used in the numerical scheme used in Section \ref{QNM-frequencies}.

It should be emphasized that this result is valid for a wide class of self-interaction potentials, which satisfy (\ref{eq:cond_V}), and of course, includes the power-law potentials.

\section{Power-law potential}\label{monomial_V}
Scalar field self-interation in the form of the power-law potential is considered:
\begin{equation}\label{SF-potential}
V(\phi)=V_0 |\phi|^{2n},
\end{equation}
where $V_0>0$ and $n>2$ (not necessarily an integer). The latter condition means that we restrict  ourselves to the case of  the massless SF with the Coulomb behavior  at spatial infinity, which requires  $n>2$  \cite{ZhdSt,SZA2}; otherwise the asymptotic condition for  $\phi$  will be different. 
Because the solution $\phi(r)$ of (\ref{equation-phi}) has no zeros \cite{ZhdSt}, we can choose the sign by setting  $\phi(r)>0$. From now on the units are set to $G=c=1$ and the constant $V_0$ can be ruled out by rescaling $r$ and $\phi$.

We are interested in isolated configurations with spatially-flat asymptotics, which satisfy the  conditions
\begin{equation}\label{flattness}
\lim_{r\to \infty} [r(e^{\alpha}-1)]= \lim_{r\to \infty} [r(e^{-\beta}-1)] =-r_g,\quad \lim_{r\to \infty} \left[r \phi(r)\right]=Q\,,
\end{equation}
where $r_g=2M$ and $M>0$ is the configuration mass, $Q$ is the ``scalar charge''.  The global and asymptotic properties of the  solutions satisfying  (\ref{flattness}) have been studied  in \cite{ZhdSt,SZA2}. For fixed $n>2$ the solution of  system (\ref{Ein_1-0}--\ref{equation-phi}) is uniquely defined by parameters $M$ and $Q$ \cite{SZA2}.   There exists the naked singularity at the center ($r=0$); the asymptotic  behaviour of the functions $\alpha(r), \beta(r)$ and $\phi(r)$ is (\ref{asymptotics_NS_phi}, \ref{asymptotics_NS}).
We use the backwards integration starting from a sufficiently large initial radius $r_{\rm in}$ to smaller $r$ up to the origin. The  ``initial" conditions at $r_{\rm in}$ were specified  using the  asymptotic relations \cite{ZhdSt,SZA2}   for $r\to\infty$, $n>2$ that can be inferred from (\ref{flattness}): 
\be
\label{asymptotics_NS-infty}
e^{\alpha}=\left(1-\frac{r_g}{r}\right)\left[1+O\left(\frac{\mu(r)}{r^2}\right)\right] ,~ e^{- \beta}= \left(1-\frac{r_g}{r}\right)\left[1+\frac{4\pi Q^2}{r^2}+O\left(\frac{\mu(r)}{r^2}\right)\right] \,,
\ee
\be
\phi(r)=\frac{Q}{r}\left[1+\frac{r_g}{2r}+\frac{n|Q|^{2n-2}}{(n-2)(2n-3)r^{2n-4}}+O\left(\frac{\mu(r)}{r^2}\right) \right]\,.
\ee
where $\mu(r)= 1/r$ for $n\ge 3$ and $\mu(r)= 1/r^{2n-4}$ for $2<n<3$.
This yields for $r\to{\infty}$
\eq{r^*= r+r_g\ln (r/r_g)+O\left(\frac{1}{r}\right),\quad W_{\rm eff}(r,l)=\frac{l(l+1)}{r^2}+O\left(\frac{1}{r^3}\right),\quad W_{\rm eff}(r^*)=\frac{l(l+1)}{r^*{}^2}\left(1+O\left(\frac{\ln r^*}{r^*}\right)\right) .}
 
\begin{figure} [h]  
\includegraphics[width=180mm]{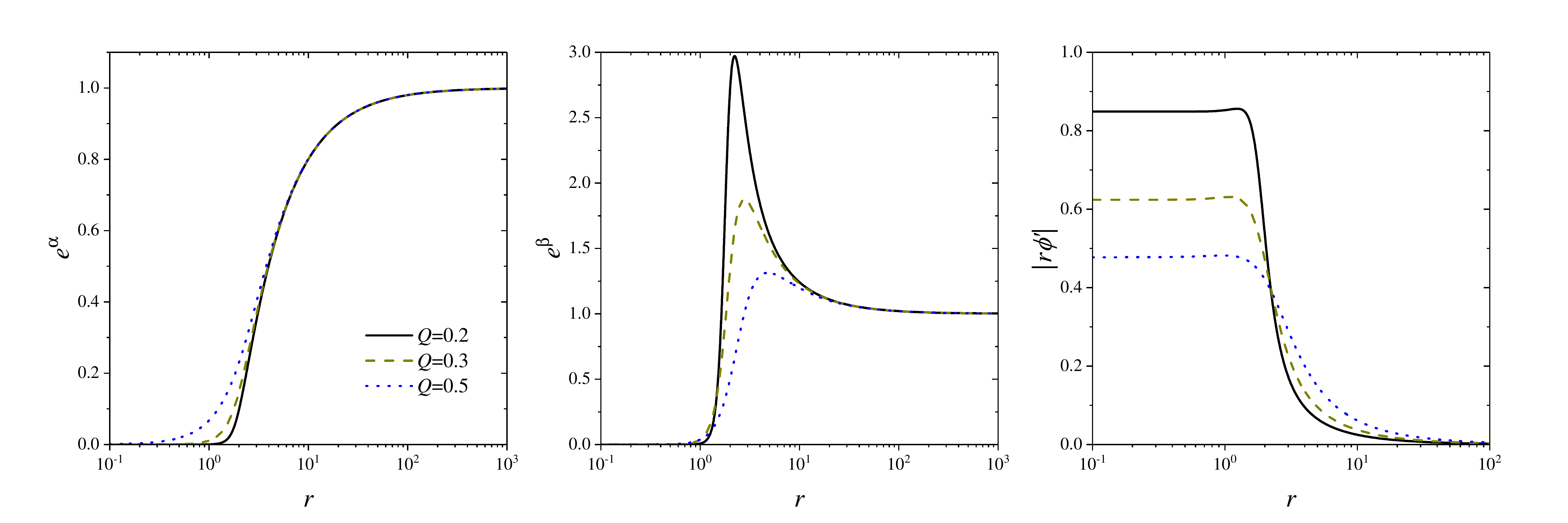}
 \caption{Typical behaviour of the background solutions of Einstein--SF equations. The examples  are given for $M=1$, $n=3$ and (a) $Q=0.5$; (b) $Q=0.3$, (c) $Q=0.2$. For smaller $Q$, the metric gets closer to the abscissa  near the origin and approach the Schwarzschild values for $r>r_g$. Right panel shows how  $|r\,d\phi/dr|$ tends to  asymptotic value $\xi$ of  (\ref{asymptotics_NS_phi}).}
      \label{BackgroundSolutions}
\end{figure}

In our numerical calculations we have considered $r_{\rm in}=10^5$. 
  Examples of the qualitative behavior of the solutions are shown in Fig. \ref{BackgroundSolutions} for some values of $Q$. The scalar field $\phi(r)$ decreases monotonically for all possible values of $M,Q,n$. As for the metric coefficients,  $e^{\alpha(r)}$ increases monotonically from zero, whereas $e^{\beta(r)}$ increases form zero to some maximum and then decreases  to $1$. For large $Q$ the maximum of $e^{\beta(r)}$ is almost invisible, but for small $Q\to 0$ the maximum of $e^\beta$  increases indefinitely. Also, for  $Q\to 0$ the metric   in the region $r>r_g$ tends to the Schwarzschild values, so all physical effects occurring  in the this region   will be indistinguishable from those   of the Schwarzschild black hole. On the other hand, in the inner region ($r<r_g$), there is a cardinal difference from the Schwarzschild black hole even for small $Q$. This gives us a motivation to look for  observational effects associated   with the inner region, that may arise in the analysis of perturbations.
The latter relation corresponds to (\ref{W_inf}).
 
 \begin{figure}[h]
 \centering
    \includegraphics[width=190mm]{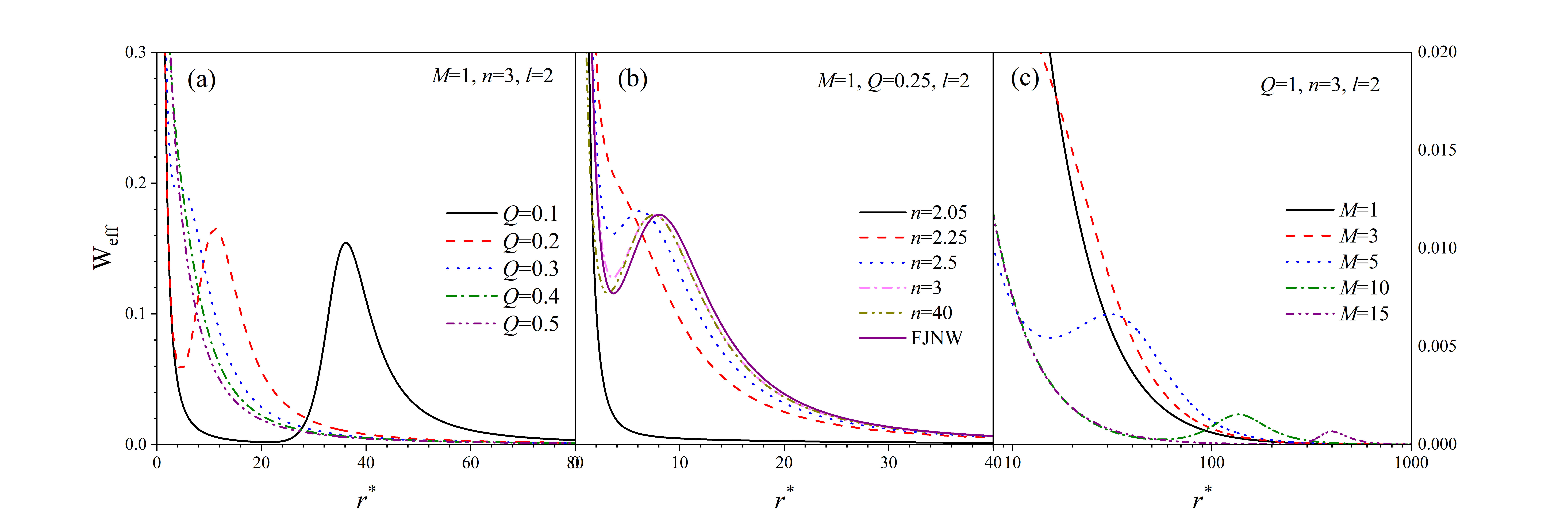}
\caption{Typical behaviour of the effective potential $W_{\rm eff}(r,l=2)$: (a)  $M=1$, $n=3$ for different Q;   (b)  $Q=0.25$, $M=1$, different $n$;   (c) $Q=1$, $n=3$, different $M$. }
      \label{Veff_example}
\end{figure} 

\section{Numerical solutions to the master equation and  frequencies of fundamental QNM} \label{QNM-frequencies}
Our aim is  to extract  the fundamental QNM frequencies\footnote{Here the fundamental  $\omega$ is that of QNM   with the least damping, that is with the minimal value $\omega_I=\text{Im}(\omega)$} of the axial  space-time perturbations governed by (\ref{master_eq}). To do this, we solve this equation keeping in mind that the QNM frequencies  reflect the internal properties of the background configuration; they do not depend on the form of the initial perturbation, which can be chosen with a large degree of arbitrariness.

The background configuration manifests itself by means of the wave potential in  (\ref{master_eq}), which depends on the configuration parameters.   
Typical behaviour of $W_{\rm eff}$ is shown in Fig. \ref{Veff_example} for different  $Q$, $M$  and $n$, in panels (a), (b), (c), respectively. For a certain set of parameters, the effective potential can have a local maximum, which can disappears with increasing $Q$ or decreasing $M$.

In order to integrate the master equation (\ref{master_eq}) and 
analyze the QNMs spectrum we use the time-domain integration method \cite{Konoplya2011}.

It is convenient to rewrite the  wave equation (\ref{master_eq}) in terms of the null coordinates  $u=t-r^*$, $v=t+r^*$: 
\eq{\label{Wave-eq_c}
4\frac{\partial^2}{\partial u \partial v}\Psi(u,v)+W_{\rm eff}(u,v,l)\Psi(u,v)=0.}
Taking into account the remarks at the end of Section \ref{General} in connection with asymptotic relations (\ref{asy-V-zero}), following \cite{Wald1980}, we impose the null  Dirichlet condition at $r^* = 0$:
\begin{equation}\label{null_Dirichlet}
\Psi(u=v,v)=0\,.
\end{equation}
Assuming that the ringtone frequencies are not sensitive to the form of the initial
perturbation 
we chose a condition on characteristic $u=0$ corresponding to the  Gaussian wave packet: 
\begin{equation}
\Psi(u=0,v)=\exp\left[-\frac{(v-v_c)^2}{2\sigma^2}\right] \,,
\end{equation}
centered at $v = v_c= 50$,  the width    $\sigma= 1$.  
\begin{figure}[h!]
    \includegraphics[width=1.0\textwidth]{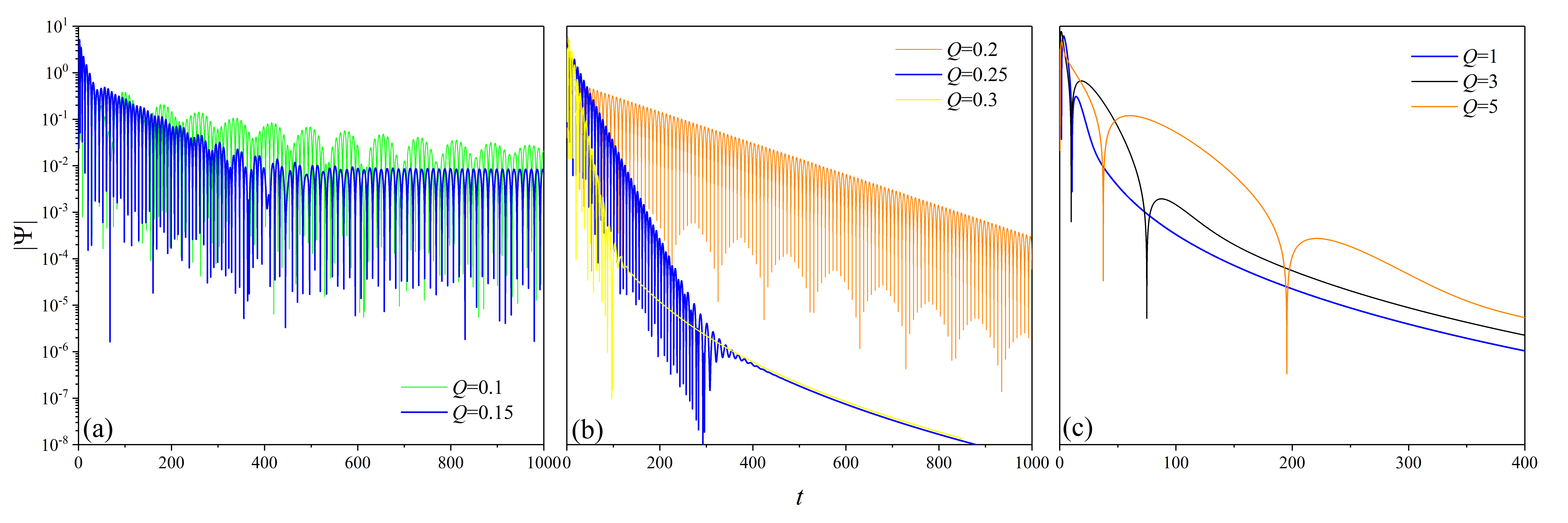}
\caption{Typical behaviour of $|\Psi(t,r^*=200)|$ for $l=2$, $n=3$, $M=1$. The left panel shows  QNM signal with echo noise  due to the existence of the potential peak. For $Q=0.15$ (blue graph) the echo dominates for $t\lesssim 500$; then the ringdown dominates. The blue and yellow curves on panel (b) show, how after the ringdown time we have typical power-law tails. The panel (c) also illustrates that for larger values of $Q$, the values of  $\omega_R$ and $\omega_I$  becomes smaller and then, after finishing oscillations we have  power-law tails. }
      \label{fig:profiles1}
\end{figure} 

 \begin{figure}[h!]
     \centering
     \includegraphics[width=85mm]{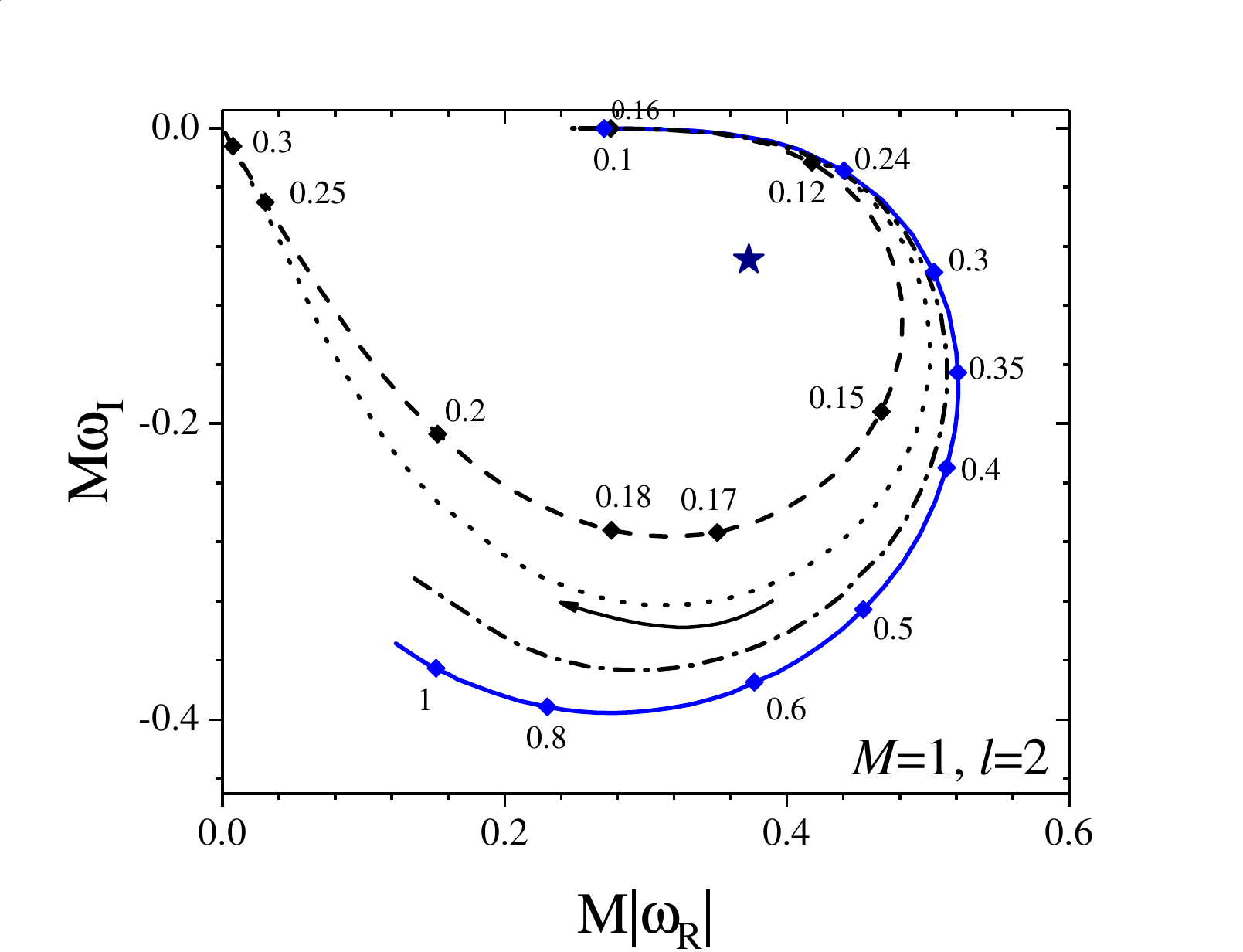}
     \includegraphics[width=85mm]{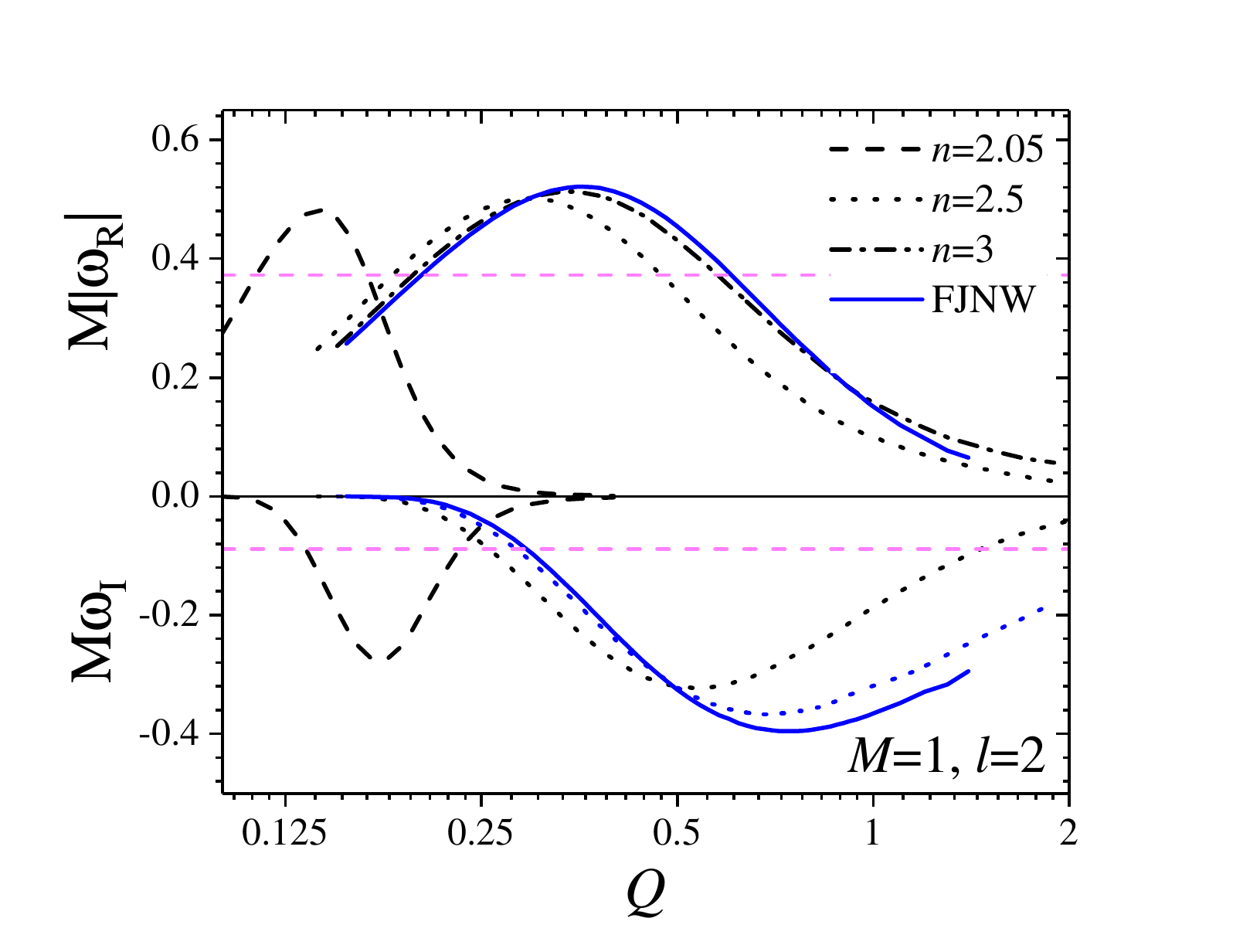}
     \caption{Fundamental frequencies $\omega$ as functions of $Q$ for three   $n$ values (indicated in the upper right corner of the right panel); $M=1$, $l=2$. Solid blue curves correspond to the FJNW case.  Left panel:  trajectories   $\omega(Q)$ in the complex plane; black arrow shows the direction of increasing $Q$ and the numbers near corresponding points of the dashed and solid blue curves indicate the values of $Q$; the blue star shows the Schwarzschild case. Right panel shows separately $\omega_R (Q)$ and $\omega_I(Q)$; the horizontal dashed lines correspond to the Schwarzschild case.}
     \label{fig_curves of Q,l=2}
 \end{figure} 
 
   \begin{figure}[h!]
     \centering
     \includegraphics[width=85mm]{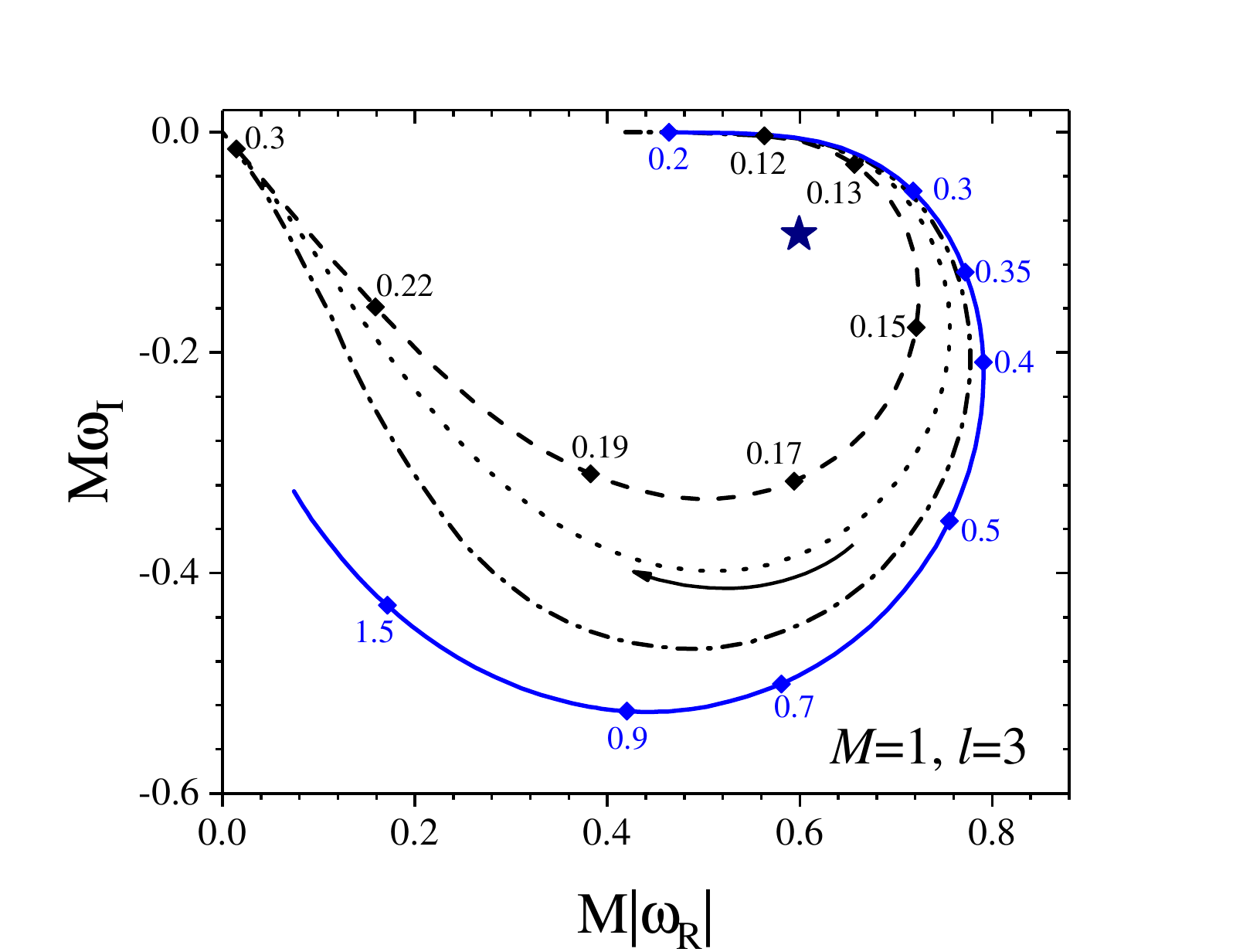}
     \includegraphics[width=85mm]{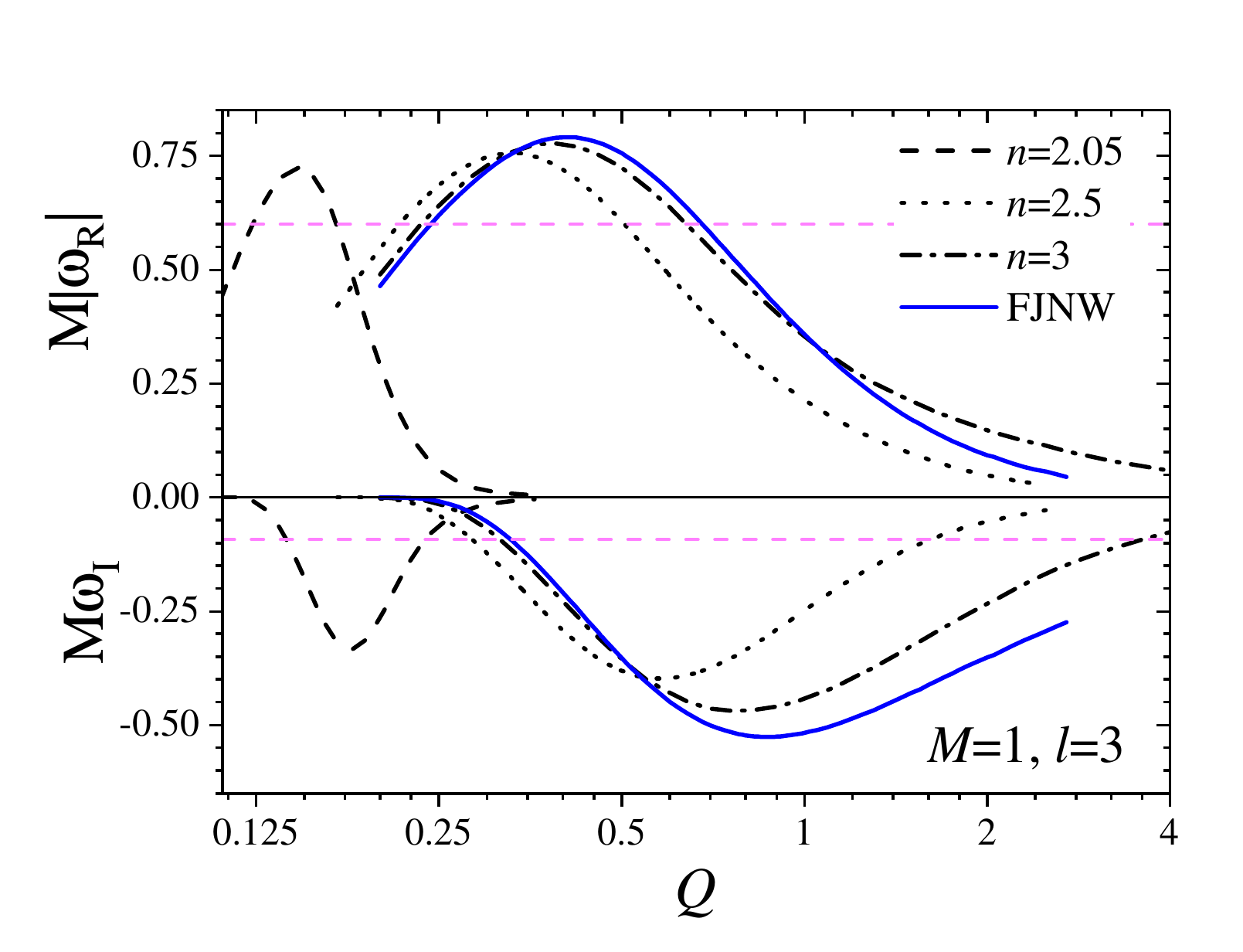}
     \caption{The same as on Fig. \ref{fig_curves of Q,l=2} for $M=1$, $l=3$.}
     \label{fig_curves of Q,l=3}
 \end{figure} 
 
  \begin{figure}[h!]
     \centering
     \includegraphics[width=85mm]{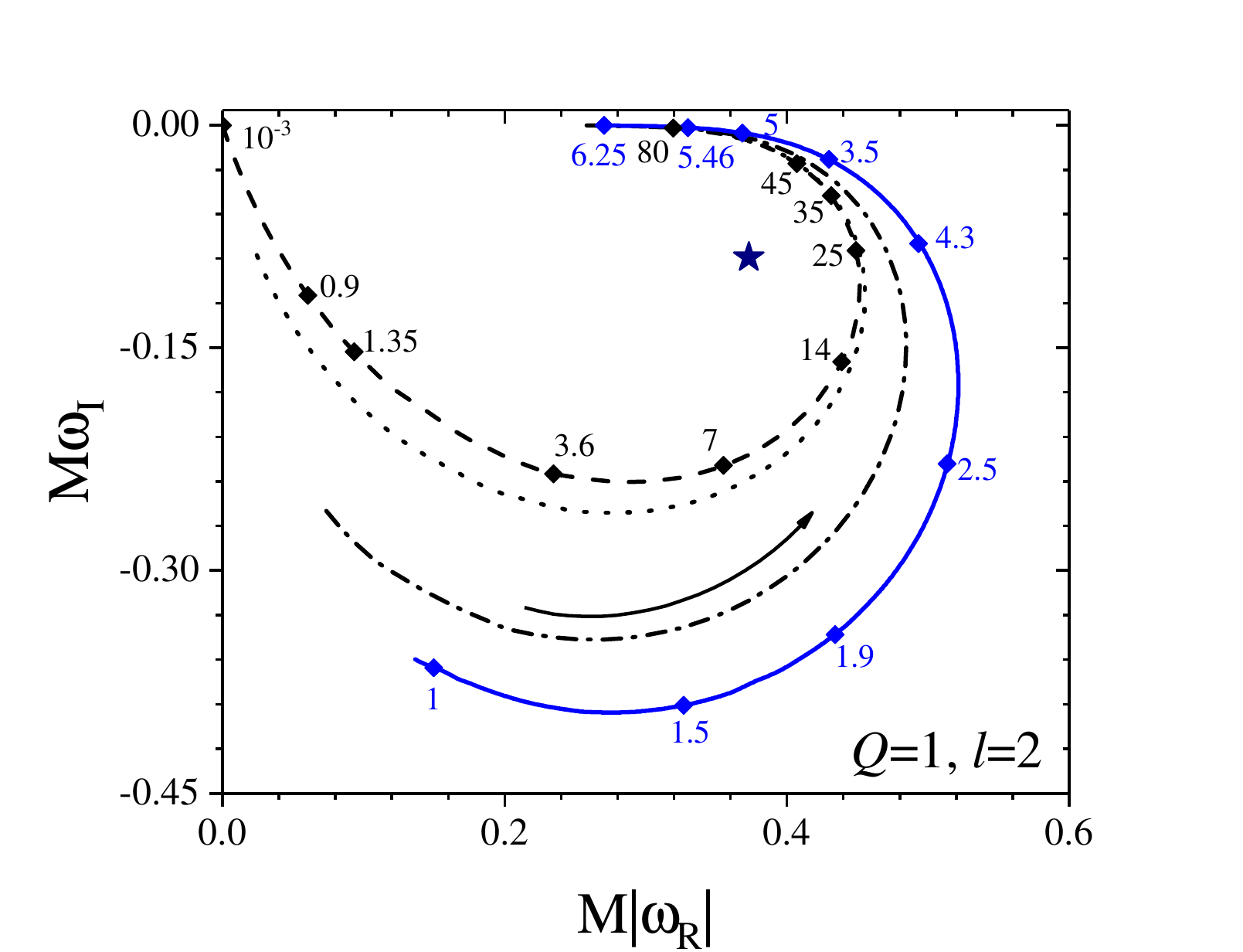}
     \includegraphics[width=85mm]{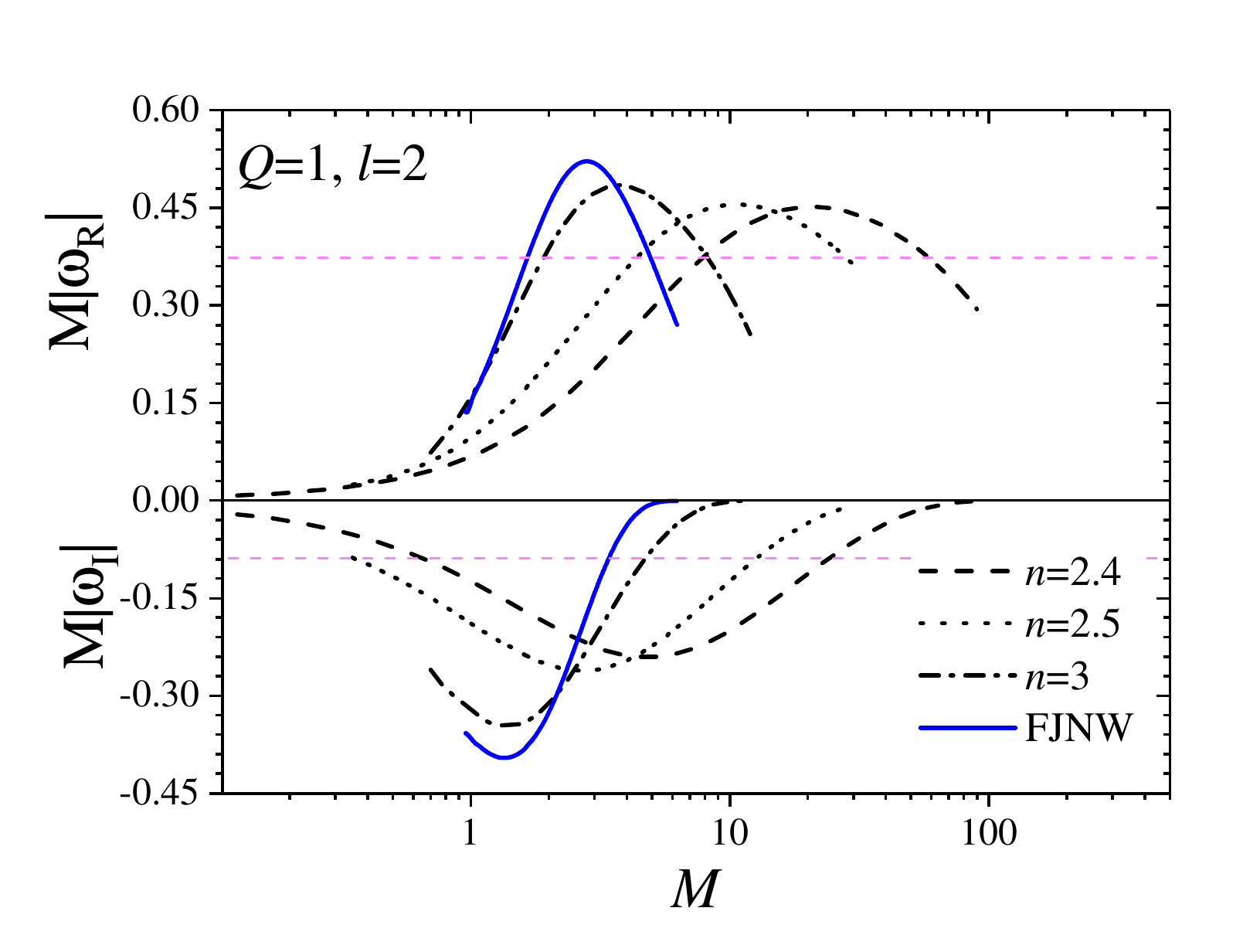}
     \caption{ Fundamental frequencies   $\omega$  as functions of $M$ for different  $n$ values (indicated in the right lower corner of the right panel); $l=2$, $Q=1$.  Left panel:  trajectories  $\omega(M)$   in the complex    plane; black arrow shows the direction of increasing $M$ and the numbers near the corresponding points of the dashed and solid blue curves indicate the values of $M$; the blue star shows the Schwarzschild case. Right panel shows separately dependencies $\omega_R (M)$ and $\omega_I(M)$; the horizontal dashed lines correspond to the Schwarzschild case.}
     \label{fig_curves of M,l=2}
 \end{figure} 
 
 \begin{figure}[h!]
     \centering
     \includegraphics[width=85mm]{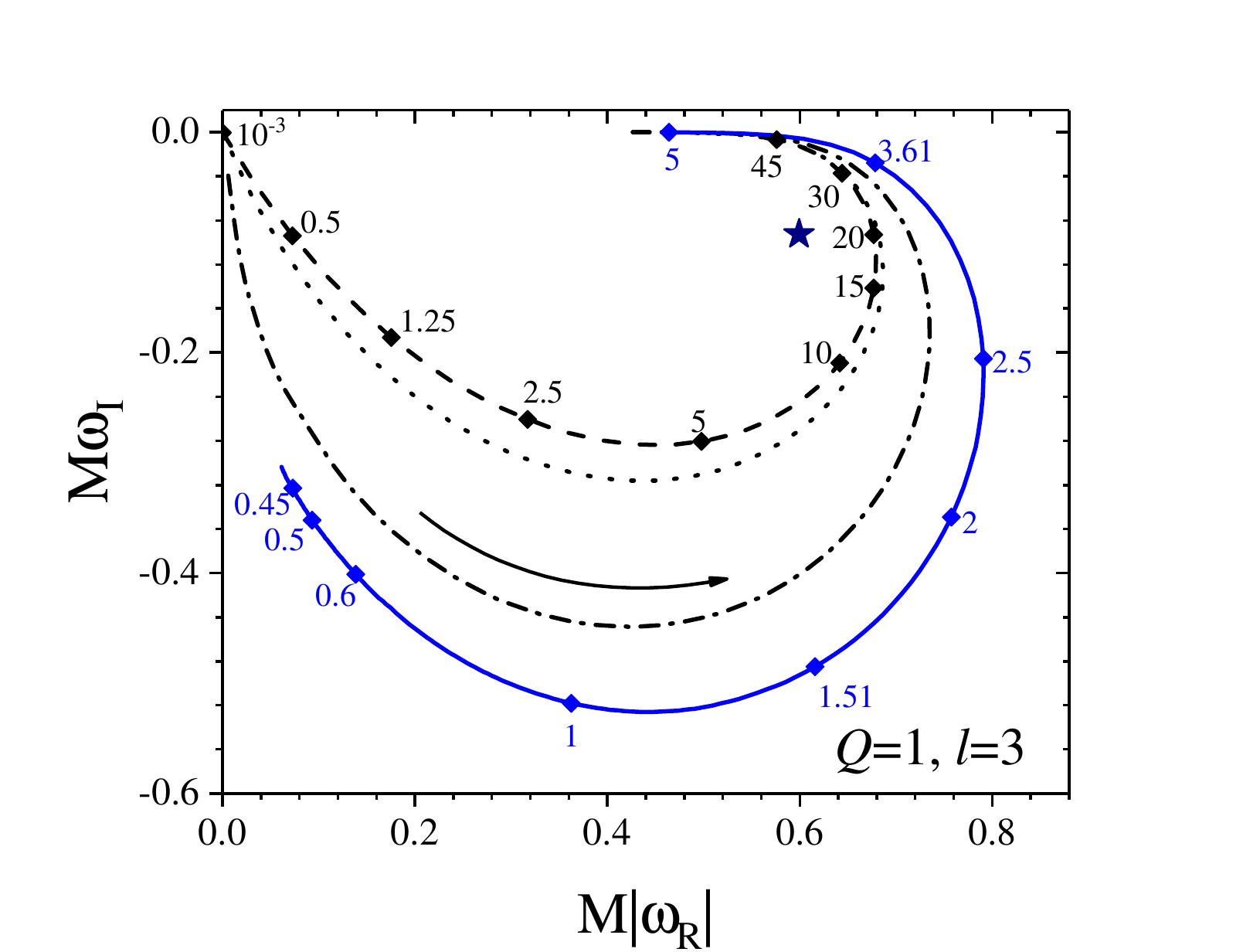}
     \includegraphics[width=85mm]{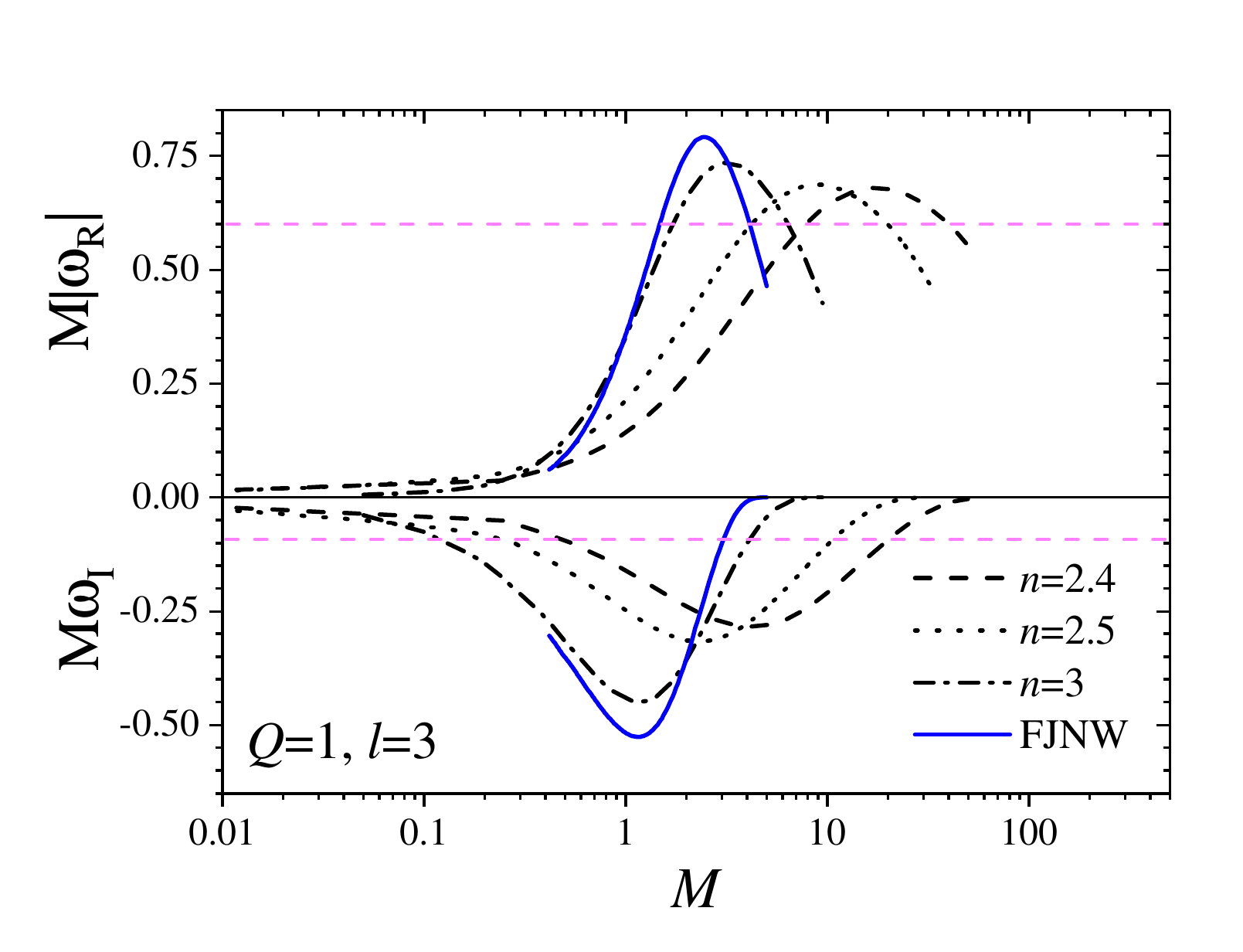}
     \caption{The same as on Fig.  \ref{fig_curves of M,l=2} for $Q=1$, $l=3$.}
     \label{fig_curves of M,l=3}
 \end{figure} 
  \begin{figure}[h!]
     \centering
     \includegraphics[width=85mm]{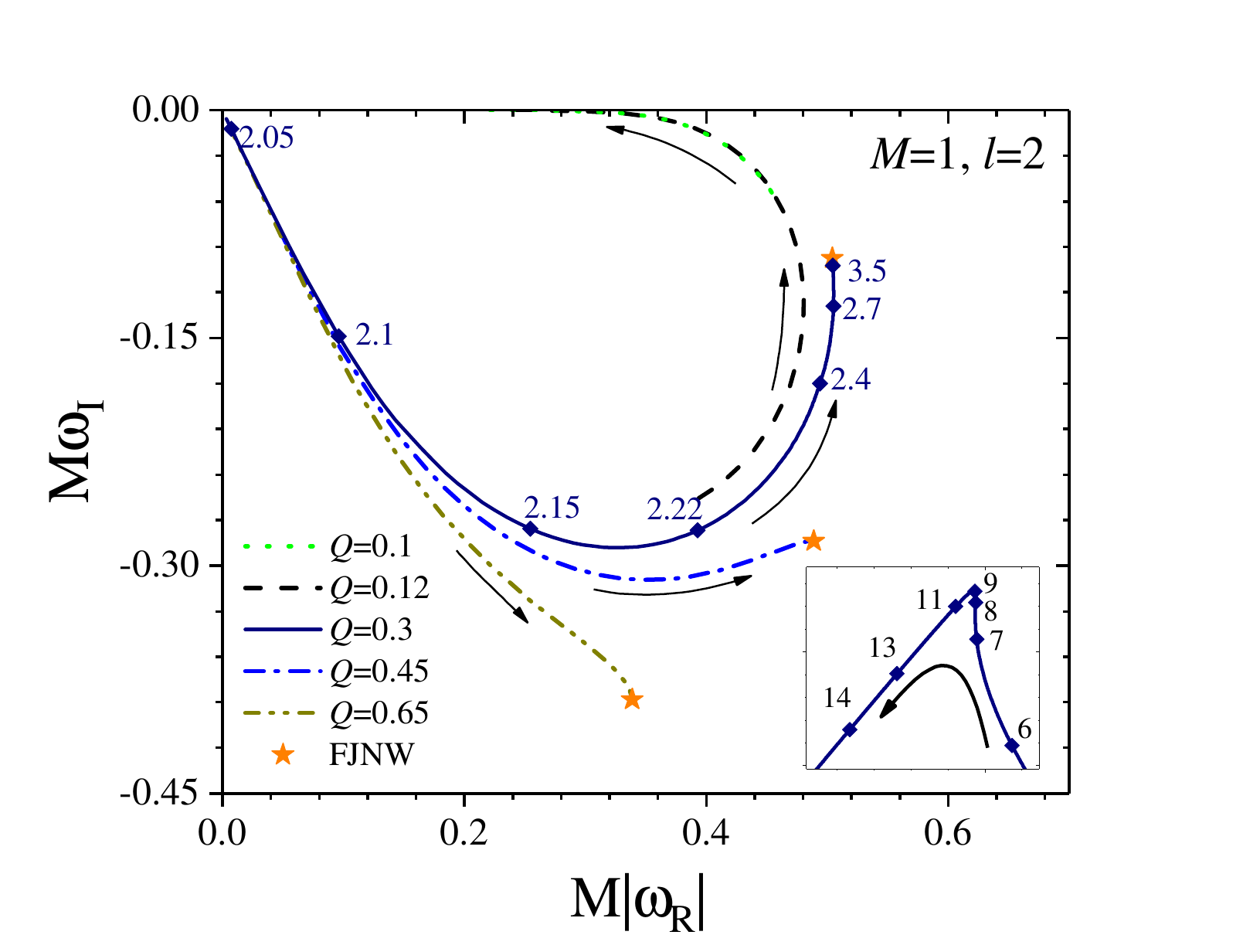}
     \includegraphics[width=85mm]{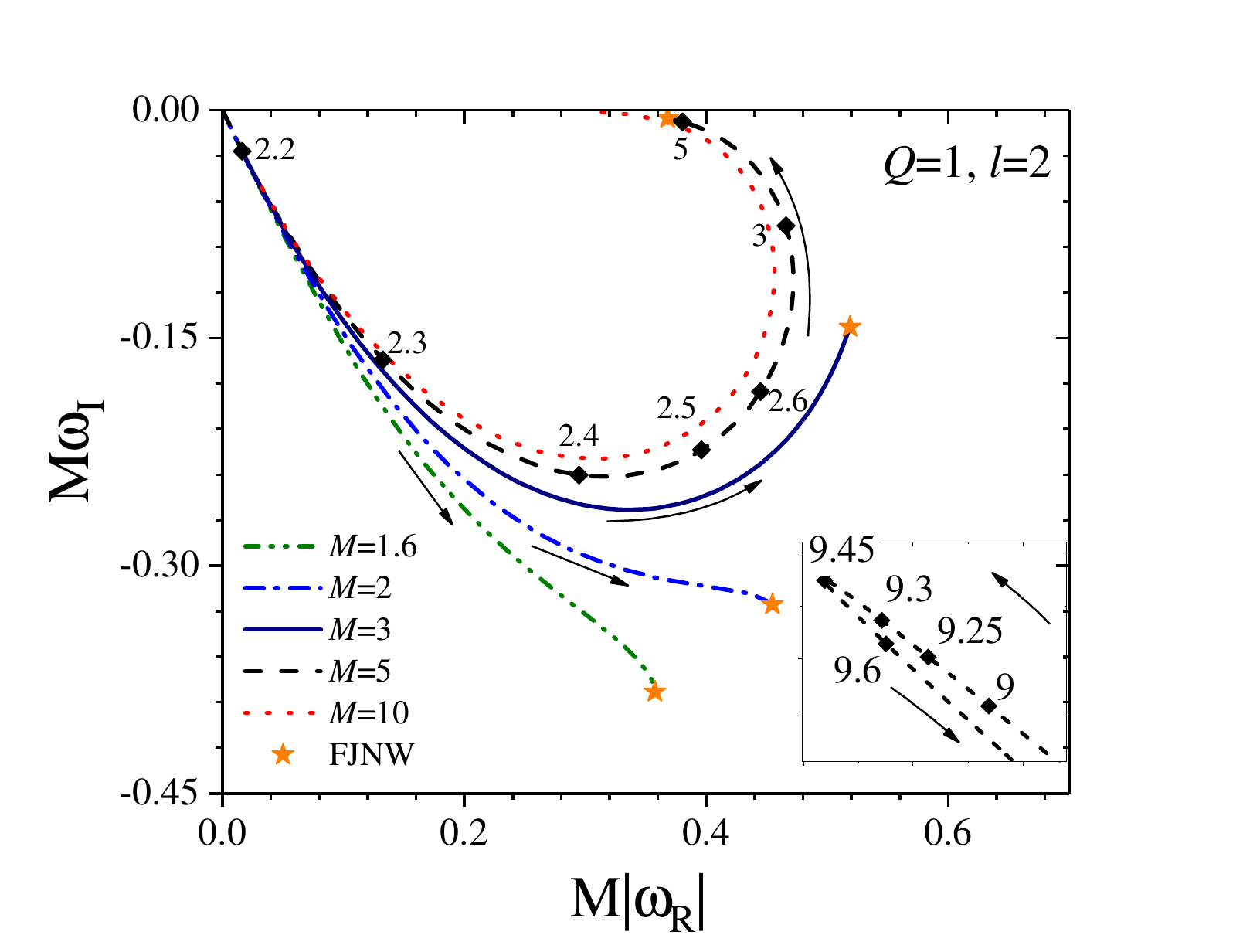}
     \caption{The fundamental frequencies $\omega$ in the complex plane $\omega_R\times\omega_I$  as functions of $n$ for $l=2$. Left panel: trajectories  $\omega(n)$ for fixed $M=1$ and several different $Q$; the numbers near the corresponding points of some curves indicate the values of $n$. Right panel: the same  for $Q=1$ and several different $M$. Black arrows show the direction of increasing $n$.  As $n$ increases, $\omega(n)$ approach FJNW values (orange stars); however, for sufficiently large $n$ we found  typical small features shown in the insets.}
     \label{fig-of-n}
 \end{figure} 
 \FloatBarrier

To obtain numerical solutions, we use a second-order discretization scheme proposed by  Chirenti and Rezzolla  \cite{Chirenti_2007}.
\eq{\Psi_N=(\Psi_W+\Psi_E)\frac{16-\Delta^2W_{\rm eff}(S)}{16+\Delta^2W_{\rm eff}(S)}-\Psi_S+O(\Delta^4),}
where the indices $(N, W, E, S)$  correspond to the points of the space-time  triangular grid and as follows: $N= (u + \Delta, v +\Delta)$, $W = (u + \Delta, v)$, $E = (u, v +\Delta )$, $S= (u, v)$;  $\Delta$ is  the spacing between the grid points.

To extract  fundamental frequencies $\omega=\omega_R+i \omega_I$, we fitted the time-domain profiles for a sufficiently large times and fixed $r^*$  by a sum of complex  exponentials $\Psi(t)\simeq\sum\limits_{j=1}^{p} A_j e^{-i\omega_j t}$ using the Prony method \cite{prony}. 

Some typical examples of numerical solutions of (\ref{Wave-eq_c}) are presented on Fig. \ref{fig:profiles1}. The left panel (a) of this figure shows  situation with small $Q$, when there is a series of echoes. The  ringdown profile of damped QNM can be retrieved at a later time when the echo contribution is negligible. The series of the ringdown oscillations is followed by a  power-law damping.

The derived fundamental QNM frequencies  are presented in Tables \ref{Table1-of-Q}--\ref{Table2-of-n}.
We plotted the trajectories of these  frequencies in the complex $\omega$-plane as functions of  one of the  parameters $M,Q,n$  with  other parameters fixed (Figs. \ref{fig_curves of Q,l=3}-- \ref{fig-of-n}).
 As can be seen from Figs. \ref{fig_curves of Q,l=2},  \ref{fig_curves of Q,l=3}, for smaller $Q$ the fundamental frequencies get closer  the FJNW values for all fixed $n$ and $M$. 
 Analogously,  Figs. \ref{fig_curves of M,l=2}, \ref{fig_curves of M,l=3}   show how $\omega$ tend to the FJNW value as $M$ grows.

As  $n$ grows, it seems that the fundamental frequencies approach  FJNW ones (see Fig. \ref{fig-of-n}). However,  for  large $n$ we found a small feature shown  in the inserts in  Fig. \ref{fig-of-n}.   To be sure that this is not an artefact of calculations, we have checked the occurrence of analogous features for different choices of the parameters.  We explain these features by an influence of the region near the singularity. In fact, the QNM are formed mainly far from the center; in this region, for large $n$ the wave potential (\ref{master_eq}) is almost zero, as in the FJNW case.  On the other hand, the effect of the region where $|\phi(r)|>1$, becomes more tangible as $n$ grows, but it is suppressed due to the null boundary condition (\ref{null_Dirichlet}); that is why these features are almost imperceptible.

\section{Discussion}\label{Conclusions}
We have studied  axial  linear perturbations against static spherically symmetric    asymptotically flat background described by the Einstein -- SF equations with the power-law SF potential. The background solutions necessarily have NS at the origin.
The characteristic behavior of the perturbations induced by an initial pulse, has three main stages: (i) possible  domination of echo-signals, (ii) the ringdown stage followed by (iii) monotonous decay. The role and duration of these stages differ  depending on the configuration parameters affecting the wave potential (\ref{positive_Veff});  for example, sometimes the stage (i) may be  substantially absent. We were mostly interested in  stage (ii) of the QNM domination yielding the fundamental frequencies of these modes.  
The results of the numerical simulations are  presented in Tables \ref{Table1-of-Q}, 
\ref{Table2-of-n}; they are used 
 in Figs. \ref{fig_curves of Q,l=2}-\ref{fig-of-n}
  for different sets of the  background configuration parameters. 

We also  draw  attention to the analytic result described by equations (\ref{master_eq}, \ref{positive_Veff}) that have been used for the general  SF potential $V(\phi)\ge 0$ to study stability issues. In particular, we used this in case of the rather general  $V(\phi)$, when the background solutions  have an appropriate asymptotic near the naked singularity. Of course, this does not yet solve the entire problem of the stability of the NS, since we have limited ourselves to axial perturbations. The polar perturbations  deserve a separate study and we plan to consider this issue elsewhere.

The presence of the non-zero SF potential manifests itself in the values $\omega$ of the  fundamental QNM frequencies that are different for different background configuration parameters. There is also a considerable difference  from the FJNW values. However,   the qualitative behavior of trajectories in the $\omega$-plane is  rather similar to the FJNW curves. For fixed $M$, the $\omega$-trajectories get closer to the  FJNW curves in the intervals of smaller $Q$ (Figs. \ref{fig_curves of Q,l=2} - \ref{fig_curves of Q,l=3}). Therefore,  for smaller SF the influence of the power-law  potential decreases. A similar trend is also observed for a fixed $Q$, when $M$ gets larger (Figs. \ref{fig_curves of M,l=2} - \ref{fig_curves of M,l=3}).
For large $n$, the wave potential and $\omega$-trajectories  can be well approximated by the FJNW curves; nevertheless the nonlinear effects manifest themselves  though the small features  that are almost  imperceptible.
On the whole, the FJNW case satisfactory describes the qualitative behavior of $\omega$-trajectories. Of course, this cannot be said about the numerical values of the fundamental QNM frequencies.

If  SF is small enough, its influence   in the outer region ($r>r_g$) is  becomes negligible. In particular, this concerns the effects in motion of particles and photons, structure of accretion disks etc. 
On the other hand, it is important to note that in the region $r<r_g$, the  significant differences from the Schwarzschild case persist  even for  small $Q$. This can be explained by different boundary conditions: the Dirichlet condition  at the naked singularity in presence of SF versus the in-going wave condition at the black hole horizon in the Schwarzschild case. The boundary condition at the origin feels the difference of NS from  the black hole even when SF is arbitrarily small. 
Thus,  the fundamental frequencies of the configuration considered,  differ qualitatively and numerically from those in case of the  Schwarzschild solution even for  a small scalar field. One can  suggest that this is a  fairly general property of static systems with the scalar field leading to the occurrence of naked singularity.

\begin{acknowledgments}
We are indebted to  Prof. Luciano~Rezzolla for helpful discussions. The work of V.I.Z. was supported by the National Research Foundation of Ukraine under Project No. 2020.02/0073.  O.S. is grateful to Igor Klebanov and Horst Stoecker for their kind hospitality.
\end{acknowledgments}

\section{{Appendix: tables}}\label{Appendix}
\begin{table}[!hbp]
\caption{The fundamental quasinormal mode $M\omega$ of the NS with nonlinear SF ($M=1$, $n=(2.1,3)$) and FJNW for various values of $Q$. The first line corresponds to the Schwarzschild black hole.}
\begin{tabular}{|c|cc|cc|cc|cl|}
\hline
& \multicolumn{2}{c|}{$n=2.1$}     & \multicolumn{2}{c|}{$n=3$}    & \multicolumn{2}{c|}{FJNW} \\ \hline
$Q$  & \multicolumn{1}{c|}{$l=2$} & $l=3$ & \multicolumn{1}{c|}{$l=2$} & $l=3$  & \multicolumn{1}{c|}{$l=2$} & $l=3$ \\ \hline
$0$   & \multicolumn{1}{c|}{$0.3730-0.0891i$}    & $0.5993-0.0927i$ & \multicolumn{1}{c|}{$0.3730-0.0891i$} & $0.5993-0.0927i$ & \multicolumn{1}{c|}{$0.3730-0.0891i$}  & $0.5993-0.0927i$ \\ \hline
$0.2$ & \multicolumn{1}{c|}{$0.4303-0.2425i$}    & $0.6917-0.2536i$ & \multicolumn{1}{c|}{$0.3815-0.0077i$} & $0.489 -0.0002i$  & \multicolumn{1}{c|}{$0.3683-0.0052i$}  & $0.4643-0.00009i$ \\ \hline
$0.25$  & \multicolumn{1}{c|}{$0.2172 -0.2558i$}    &$0.4005-0.322i$  & \multicolumn{1}{c|}{$0.4639-0.048i$} & $0.6413-0.01445i$  & \multicolumn{1}{c|}{$0.45481-0.0381i$}  & $0.6196-0.0084i$  \\ \hline
$0.3$  & \multicolumn{1}{c|}{$0.0874 -0.1371i$}    & $0.1731-0.1748i$ & \multicolumn{1}{c|}{$0.5051-0.112i$} & $0.7293-0.0701i$  & \multicolumn{1}{c|}{$0.5042	-0.09739i$}  & $0.7179-0.0534i$ \\ \hline
$0.35$ & \multicolumn{1}{c|}{$0.0346 -0.0573i$}    & $0.0692-0.0712i$ & \multicolumn{1}{c|}{$0.5131-0.1799i$} &  $0.7703-0.1484i$  & \multicolumn{1}{c|}{$0.5214-0.166i$}  & $0.7723-0.1268i$ \\ \hline
$0.45$  & \multicolumn{1}{c|}{$0.0066 -0.0112i$}    & $0.01323 -0.0137i$  & \multicolumn{1}{c|}{$0.4682-0.2874i$} & $0.7567-0.2992i$   & \multicolumn{1}{c|}{$0.4888-0.2839i$}  & $0.7824-0.286i$ \\ \hline
$0.55$  & \multicolumn{1}{c|}{$0.00194 -0.0034i$}    & $0.0039 -0.0041i$ & \multicolumn{1}{c|}{$0.3933-0.3449i$} & $0.6803-0.3998i$  & \multicolumn{1}{c|}{$0.4161-0.3558i$}  & $0.7172-0.4067i$ \\ \hline
$0.65$  & \multicolumn{1}{c|}{$0.00081 -0.0014i$}    & $0.0016 -0.0017i$ & \multicolumn{1}{c|}{$0.3203-0.3653i$} & $0.5898-0.4512i$   &  \multicolumn{1}{c|}{$0.3388-0.388i$}   & $0.6269-0.4791i$ \\ \hline
\end{tabular}
\label{Table1-of-Q}
\end{table}
 \begin{table}[h]
\caption{The fundamental quasinormal mode $M\omega$ of the NS with nonlinear SF ($M=1$, $Q=(0.15,~0.3,~0.45)$) and FJNW for various values of $n$. The first and second lines correspond to the Schwarzschild black hole and FJNW solutions, respectively.
}
\begin{tabular}{|c|cc|cc|cc|cl|}
\hline
& \multicolumn{2}{c|}{$Q=0.15$}     & \multicolumn{2}{c|}{$Q=0.3$}    & \multicolumn{2}{c|}{$Q=0.45$} \\ \hline
$n$  & \multicolumn{1}{c|}{$l=2$} & $l=3$ & \multicolumn{1}{c|}{$l=2$} & $l=3$  & \multicolumn{1}{c|}{$l=2$} & $l=3$ \\ \hline
$Schw$   & \multicolumn{1}{c|}{$0.3730-0.0891i$}    & $0.5993-0.0927i$ & \multicolumn{1}{c|}{$0.3730-0.0891i$} & $0.5993-0.0927i$ & \multicolumn{1}{c|}{$0.3730-0.0891i$}  & $0.5993-0.0927i$ \\ \hline
FJNW  & \multicolumn{1}{c|}{$0.2436 - 0.00006i$}    &$0.2796 -6\cdot10^{-8}i$ & \multicolumn{1}{c|}{$0.5042-0.0974i$} & $0.7183-0.0535i$  & \multicolumn{1}{c|}{$0.4888-0.2839i$}  & $0.7824-0.286i$  \\ \hline
2.05  & \multicolumn{1}{c|}{$0.467-0.1918i$}    & $0.721-0.1773i$ & \multicolumn{1}{c|}{$0.0073-0.0123i$} & $0.0146-0.0151i$ & \multicolumn{1}{c|}{$0.0006 -0.001i$}  & $0.0012 -0.0013i$ \\ \hline
2.1 & \multicolumn{1}{c|}{$0.4517-0.0484i$}    & $0.6305 -0.0158i$  & \multicolumn{1}{c|}{$0.0876-0.1369i$} & $0.173-0.1748i$  & \multicolumn{1}{c|}{$0.0066 -0.0112i$}  & $0.01324 -0.0137i$ \\ \hline
2.3  & \multicolumn{1}{c|}{$0.3178-0.0012i$}    &  $0.3861 -6\cdot10^{-6}i$ & \multicolumn{1}{c|}{$0.4698-0.2222i$} & $0.7352 -0.2153i$   & \multicolumn{1}{c|}{$0.2654 -0.2943i$}  & $0.4839 -0.3661i$ \\ \hline
2.5  & \multicolumn{1}{c|}{$0.2802 -0.00029i$}    &  $0.33 -6\cdot10^{-7}i$ & \multicolumn{1}{c|}{$0.5016 -0.1558i$} & $0.7475 -0.1226i$   & \multicolumn{1}{c|}{$0.3952 -0.306i$}  & $0.6675 -0.3476i$ \\ \hline
3 & \multicolumn{1}{c|}{$0.2532-0.00009i$}    & $-$ & \multicolumn{1}{c|}{$0.5052-0.112i$} & $0.7293 -0.0701i$  & \multicolumn{1}{c|}{$0.4682-0.2874i$}  & $0.7567 -0.2992i$ \\ \hline
4 & \multicolumn{1}{c|}{$0.245-0.00007i$}    & $-$ & \multicolumn{1}{c|}{$0.5044-0.0995i$} & $0.7201 -0.056i$  & \multicolumn{1}{c|}{$0.4867 -0.2837i$}  &  $0.7793 -0.2869i$\\ \hline
5 & \multicolumn{1}{c|}{$0.2439	-0.0000633i$}    & $-$ & \multicolumn{1}{c|}{$0.50427-0.0979i$} &  $0.7188 -0.0541i$ & \multicolumn{1}{c|}{$0.4885 -0.2839i$}  & $0.7819 -0.2862i$ \\ \hline
7 & \multicolumn{1}{c|}{$0.2436 -0.0000625i$}    & $-$ & \multicolumn{1}{c|}{$0.50426-0.0976i$} &  $0.7184 -0.0536i$ & \multicolumn{1}{c|}{$0.48882 -0.28394i$}  & $0.7824 -0.2861i$ \\ \hline
10 & \multicolumn{1}{c|}{$0.2436-0.0000624i$}    & $-$ & \multicolumn{1}{c|}{$0.50426-0.0975i$} &  $0.7184-0.0535i$   & \multicolumn{1}{c|}{$0.48886 -0.28395i$}  & $0.7825 -0.2861i$ \\ \hline
\end{tabular}
\label{Table2-of-n}
\end{table}

\bibliography{references.bib}
\end{document}